\documentstyle[12pt,epsf]{article}

\begin{document}

\title{CAN MAGNETIC FIELDS OF ASTROPHYSICAL OBJECTS BE FUNDAMENTAL?}
\author
{Tonatiuh Matos\thanks{Permanent Adress: Dpto. de F\'{\i}sica, CINVESTAV, PO. Box 
14-740, M\'exico 07000, D.F., M\'exico. E-mail: tmatos@fis.cinvestav.mx},\\ 
Instituto de F\'{\i}sica y Matem\'aticas,\\
Universidad Michoacana de San Nicolas de Hidalgo,\\
PO. Box 2-82, 58040 Morelia, Michoac\'an, M\'exico}

\date{ \today}
%
\maketitle

\begin{abstract}
We analyze a new class of static exact solutions of Einstein-Maxwell-Dilaton 
gravity with arbitrary scalar coupling constant $\alpha$, representing a 
gravitational body endowed with electromagnetic dipole moment. This class 
possesses mass, dipole and scalar charge parameters. A discussion of the geodesic 
motion shows that the scalar field interaction is so weak that it cannot be 
measured in gravitational fields like the sun, but it could perhaps be detected in 
gravitational fields like pulsars. The scalar force can be attractive or repulsive.
 This gives rise to the hypothesis that the magnetic field of some astrophysical objects could be fundamental.
\end{abstract}


%

\newpage

A great amount of astrophysical objects in Cosmos are gravitational bodies with
magnetic dipole fields. One should suppose that the Einstein-Maxwell (EM)
theory predicts the existence of gravitational objects endowed with magnetic
dipoles. In fact there is a set of exact solutions of the EM equations
representing exterior fields of gravitational objects endowed with magnetic
dipoles \cite{kra}. Some of them are reasonably small, but they do not have the
right behavior of the gravitational field far away from the sources; the other
ones are acceptable in their behavior at infinity \cite{mank}, but the number
of terms of them is so enormous that it makes them unmanageable. On the other
hand, EM theory actually is not an unification theory, but rather a
superposition one, Einstein plus Maxwell. Here the electromagnetic field
appears as energy-momentum tensor, there is in fact no explanation of its
existence, the electromagnetic field appears like a model. For other theories
like Kaluza-Klein (KK) and Low Energy Superstring (LESS) theories, the
electromagnetic field is a component of a more general field, the existence of
gravitation and electromagnetism follows from its decomposition. In these
theories the electromagnetic field is a consequence of a more general unified
field, it is not a model. In \cite{ma25} and \cite{ma36} it is shown that the
existence of electromagnetic dipoles is natural for LESS and KK but not so
natural for EM. A class of solutions given in \cite{ma25} possesses a
gravitational field with the behavior of the Schwarzschild solution coupled
with a magnetic dipole. They are reasonably small, but they possess a scalar
field interaction, the so-called dilaton. Of course we have not observed
astrophysical objects with a scalar field interaction, but its prediction in KK
and LESS theories should be established at classical level if such theories
could be taken as realistic. In fact there are enough classical objects in
nature with manifest gravitational-electromagnetical interactions. KK and LESS
predict the existence of the dilaton at this level. In this work we will show
that the dilaton interaction cannot be measured in weak gravitational fields
like the sun, even if the sun would posses one, but it will be perhaps possible
to measure it in stronger gravitational fields like a pulsar. According to KK
and LESS theories, since these solutions posses a magnetic dipole moment
parameter and a newtonian behavior at infinity, this gives rise to the
hypothesis that the magnetic field of some astrophysical objects could be of
fundamental origin, {\it i.e.}, the magnetic field could be a consequence of a
more general scalar-gravito-electromagnetic field. In a previous work
\cite{ma25} we presented a method for finding exact solutions of the KK field
equations. These solutions represent exterior fields of a gravitational body,
endowed with arbitrary electromagnetic fields like monopoles, dipoles, etc. or
the superposition of them, from the five-dimensional point of view. Here there
exists a coupling between the electromagnetic and a scalar field, parametrized
by a coupling constant $\alpha^2=3$. In other work \cite{ma36} we generalized
this method for arbitrary $\alpha$ in order to incorporate all the most
important theories unifying gravitation and electromagnetism; KK, LESS and EM. 
The solutions found in \cite{ma36} are naked singularities in $r=2m$ for
$\alpha\not=0$. Nevertheless, we will investigate them for $r>>2m$ supposing
that they represent the exterior field of some astrophysical body. 

In the present work we analyze explicit solutions of the field equations of the
Lagrangian

\begin{equation}
{\cal L}=\sqrt{-g}[-R+2(\nabla\Phi)^2+e^{-2\alpha\Phi}F^2]
\label{lagran}
\end{equation}
obtained by this method. In (\ref{lagran}) $R$ is the curvature scalar, $F$ is the Faraday tensor, and $\Phi$ is the scalar field, the dilaton. The coupling between the dilaton and the electromagnetic field is parametrized by $\alpha$. If $\alpha=0$, 
(\ref{lagran}) is the Lagrangian of the EM theory, if $\alpha=1$, (\ref{lagran}) is the Lagrangian of the LESS theory and for $\alpha^2=3$, (\ref{lagran}) is the Lagrangian of the KK theory. 

The class of solutions we want to deal with in this work, written in
Boyer-Lindquist coordinates, reads \cite{ma25,ma36}

\[
ds^2=e^{2(k_s+k_e)}\hbox{g}^\gamma{dr^2\over 1-{2m\over r}} +
\hbox{g}^\gamma\ r^2(e^{2(k_s+k_e)}d\theta^2 + \sin^2\theta\ d\varphi^2)-
{1-{2m\over r}\over \hbox{g}^\gamma}\ dt^2
\]
\begin{equation}
A_{3,\zeta}=Q\rho\tau_{,\zeta}\ \ ,\ \ \ \ 
A_{3,\bar {\zeta}}=-Q\rho\tau_{,\bar {\zeta}}\ \ ,\ \ \ \ 
e^{-2\alpha\Phi}={\kappa^2_0e^{\tau_0\tau}\over(1-{2m\over r})
\hbox{g}^{\beta}}
\label{sol}
\end{equation}
This class of solutions can be divided into two subclasses, the subclass a)
\[
\hbox{g}=a_1e^{q_1\tau}+a_2e^{q_2\tau}, 
\]
\[
k_{s,\zeta}={\rho\over 2\alpha^2}(\lambda_{,\zeta}-\tau_0\tau_{,\zeta})^2,\ \ \ k_{e,\zeta}=-\rho\gamma q_1q_2(\tau_\zeta)^2,\ \ \ \tau_0=q_1+q_2,
\]
and the subclass b)
\[
\hbox{g}=a_1\tau+1, \ \ \ \ k_{s,\zeta}={\rho\over 2\alpha^2}(\lambda_{,\zeta})^2,\ \ \ k_e=0,\ \ \ \tau_0=0,
\]
\noindent where $\zeta=\rho+i\ z=\sqrt{r^2-2mr}\ \sin\theta+i\
(r-m)\cos\theta$. {\bf A}$=A_idx^i,\ i=1...4$ is the electromagnetic four
potential, $m$ the mass parameter, $\gamma={2\over 1+\alpha^2}$,
$\beta={2\alpha^2\over 1+\alpha^2}$; $Q,\ a_1+a_2=1,\ q_1$ and
$q_2$ are constants subjected to the restrictions
\[
2\gamma a_1a_2(q_1-q_2)^2+\kappa_0^2Q^2=0
\]
for the subclass a), and 
\[
2\gamma a_1^2-\kappa_0^2Q^2=0
\]
for the subclass b). The class of solutions (\ref{sol}) can be interpreted as a
magnetized Schwarzschild solution in dilaton gravity for $\alpha\neq 0$, for
$\alpha=0$ the construction of dipoles is different and the form of the metric
is not more similar to the Schwarzschild solution \cite{ma36}. In the following
we will assume $\alpha\neq 0$. $\lambda$ and $\tau$ are harmonic maps in a two
dimensional flat space, that means, they are solutions of the Laplace equation
\begin{equation}
(\rho\lambda_{,\zeta})_{,\bar\zeta}+(\rho\lambda_{,\bar\zeta})_{,\zeta}=0,\ \ \ \
(\rho\tau_{,\zeta})_{,\bar\zeta}+(\rho\tau_{,\bar\zeta})_{,\zeta}=0
\label{lapl}
\end{equation}
In this work we have fixed $\lambda=\ln(1-{2m\over r})$. $\tau$ determines
uniquely the electromagnetic potential. Two examples are the magnetic monopole
\begin{equation}
\tau = \ln (1-{2m\over r}), \;\;\;\;\ \ \ \ \ 
A_{3}=2mQ(1-\cos \theta ), 
\label{mon}
\end{equation}
\noindent and the magnetic dipole
\begin{equation}
\tau ={\cos\theta\over(r-m)^2-m^2\cos^2\theta} ,\;\;\;\;\ \ 
A_3={Q(r-m)\sin^2\theta\over(r-m)^2-m^2\cos^2\theta }\ \ .
\label{dip}
\end{equation}
Nevertheless, we can substitute an arbitrary electromagnetic field in
(\ref{sol}); (\ref{mon}) and (\ref{dip}) correspond to the two first spherical
harmonics, solutions of the Laplace equation (\ref{lapl}).  If $\tau=0$,
(\ref{sol}) reduces to the Schwarzschild space time coupled to the scalar field
$\Phi$, which is manifested only through $k_s$. We interprete the function g as
the contribution of the electromagnetic field to the curvature of the space
time.

If $\hbox{g}\rightarrow 1,$ and $ k_s+k_e\rightarrow 0$ for $r\rightarrow
\infty $, the solutions are asymptotically flat, and they are flat for $m=Q=0$,
at least for the examples given in (\ref{mon}) and (\ref{dip}). A general study
of the solutions contained in (\ref{sol}) will be given elsewhere \cite{mari}.

In this work we are interested in extracting some physics from dilaton
theories. In order to do so, we study the geodesic motion of test particles
traveling around the space time (\ref{sol}). Since $e^{2(k_s+k_e)}-1\sim
10^{-11}$ for a star like the sun,metric (\ref{sol}) is spherically symmetric
in this aproximation. We start from the Lagrangian

\begin{equation}
{\cal L} = e^{2(k_s+k_e)}\hbox{g}^{\gamma}{({dr\over ds})^2\over 1 - {2m\over r}} +
\hbox{g}^{\gamma}r^2~~({d\varphi
\over ds})^2 - {1-{2m\over r}\over \hbox{g}^{\gamma}}({dt\over ds})^2
\label{lgeo}
\end{equation}

\noindent where $s$ is the proper time of the test particle. In (\ref{lgeo}) we
have set $\theta={\pi\over 2}$, in this case the function $\tau$ for the dipole
field does not contribute and g$=1$. But in general, for any value of $\theta$,
the function g changes only very near to the Schwarzschild radius $r_s=2m$, but
it tends very rapidly to one far away from $r_s$, for any value of $\theta$. In
any case, in the following we will set g in all the equations where it appears.
Following any standard text book on gravitation, we first write the motion
equations. We have two constants of motion,

\[
{\delta {\cal L} \over \delta t}=0 \Rightarrow {1-{2m\over r}\over \hbox{g}^{\gamma}}({dt\over ds}) = A~~
\]
\[
{\delta {\cal L} \over \delta\varphi}=0 \Rightarrow~~\hbox{g}^{\gamma}
r^2 {d\varphi\over ds} = B~~
\]

\noindent so ${dt\over ds}$ and ${d\varphi\over ds}$ can be put in terms of $A$
and $B$. Using the equation of motion

\[
P_{\mu}P^{\mu} = -c^{2}
\]
\noindent one obtains 
\begin{equation}
-\epsilon = e^{2(k_e+k_s)}\hbox{g}^{\gamma}
{({dr\over ds})^2\over 1 - {2m\over r}} +
\hbox{g}^{\gamma}r^2~~({d\varphi
\over ds})^2 - {1-{2m\over r}\over \hbox{g}^{\gamma}}({dt\over ds})^2~~
\label{fea}
\end{equation}

\noindent where $\epsilon = c^{2}, 0, -~c^{2}$ for particles, photons and
tachyons respectively. We rewrite equation (\ref{fea}) in the more familiar
form

\begin{equation}
({dr\over ds})^2 + {e^{-2(k_s+k_e)}\over \hbox{g}^{\gamma}} 
[{B^2 \over r^{2}\hbox{g}^{\gamma}} + \epsilon]( 1 - {2m\over r}) = 
e^{-2(k_s+k_e)} A^2~~.
\label{geo}
\end{equation}

Here we have separated the part of the motion equation related with the
constant $B$ from the part related with the constant $A$ obtained from the
variation with respect to the coordinate $t$. Let us define an effective
potential by

\begin{equation}
V_{eff} = {e^{-2(k_s+k_e)}\over 2\ \hbox{g}^{\gamma}} 
[{B^2 \over r^{2}\hbox{g}^{\gamma}} + \epsilon](1 - {2m\over r})
\label{veff}
\end{equation}
\noindent and an effective energy by
\begin{equation}
E_{eff} = {1\over 2}e^{-2(k_e+k_s)} A^2~~~~.
\label{eff}
\end{equation}
\noindent in order to obtain the familiar form for the motion equation
\[
{1\over 2}({dr\over ds})^{2} + V_{eff} = E_{eff}~~.
\]

This interpretation is suggested by performing a series expansion for $r>>2m$.
Nevertheless, definitions (\ref{veff}) and (\ref{eff}) have not necessarily a
physical meaning in general. 


\begin{figure}[h]
\centerline{ \epsfysize=5cm \epsfbox{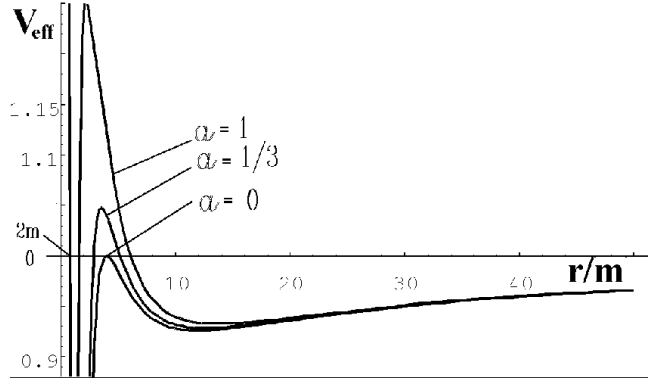}}
\caption{The effective potentials for the magnetized Schwarzschild solutions for
the KK ($a={1\over 3}$), LESS ($a=1$) theories and the Schwarzschild solution 
($a=0$). The plot is made in $m$ unities on the horizontal axis, with $l={B\over
m}=4$.}
\end{figure}


In the following we will take only the subcase b) of (\ref{sol}), here the
function $k_e=0$ and the constant $\tau_0=0$ as well. If $\theta={\pi\over2}$,
the effective potential $V_{eff}$ and the effective energy $E_{eff}$ reduce to

\[
V_{eff}=\left({(1-{m\over r})^2\over 1-{2m\over r}}\right)^a({\epsilon\over 2}- {m\epsilon\over r}+{B^2\over 2 r^2} -{mB^2\over r^3})
\]
\[
E_{eff}=\left({(1-{m\over r})^2\over 1-{2m\over r}}\right)^a~{A^2\over 2}
\]

\noindent where $a=0$ for the Schwarzschild space time and $a={1\over\alpha^2}$
for the dilatonic case. We interprete the factor $\left({1-{2m\over r}\over
(1-{m\over r})^2}\right)^a$ as the contribution of the dilaton field to the
effective potential $V_{eff}$ and to the effective energy $E_{eff}$, and the
function g as the contribution of the electromagnetic field. In figure 1 we
have plotted the effective potential for the different theories. The
qualitative behavior is very similar in all of them. In figure 2 we see the
effective energy for the same values of $\alpha$, the behavior is here very
violent; not so far away from the Schwarzschild radius, the effective energy is
constant. 


\begin{figure}[h]
\centerline{ \epsfysize=5cm \epsfbox{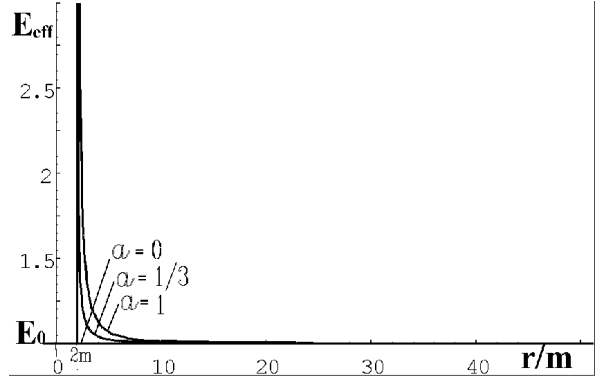}}
\caption{The effective energy for the magnetized Schwarzschild solutions.
The plot is made in $m$ unities on the horizontal axis and in $E_0$ 
unities on the vertical axis.}
\end{figure}


In order to obtain the trajectories of a test particle travelling around of a
star of sun's size, we make the standard transformation $u(\varphi)={1\over
r(\varphi(s))}$. The geodesic equation (\ref{geo}) transforms into

\begin{equation}
B^2(u')^2+\left({(1-mu)^2\over 1-2mu}\right)^a[(1-2mu)(B^2u^2+\epsilon)-A^2]=0,
\label{geou}
\end{equation}

\noindent where a prime means derivative with respect to $\varphi$. This is a
first order differential equation of the form

\begin{equation}
{1\over 2}(u')^2+V(u)=0
\label{snl}
\end{equation}

\noindent which define naturally the function $V(u)$. After derivation with
respect to $\varphi$, equation (\ref{snl}) transforms into a equation of the
form $u''+\partial_u V(u)=0$. This differential equation is very difficult to
solve and we will not try to solve it here. But for a trajectory around a star
like the sun, the mass parameter $m\sim\ 1.5$ Km., while $r\sim 10^6$ Km.,
therefore $u^3\sim 0$ is a good approximation, conserving the rest of the
terms. In that case the geodesic equation transforms into

\begin{equation}
u''+\omega^2u={m\epsilon\over B^2}+3mKu^2
\label{geope}
\end{equation}

\noindent where $\omega=\sqrt{1-{am^2\over B^2}(A^2-\epsilon)}$ and
$K=1+{am^2A^2\over B^2}$. The difference with the Schwarzschild geodesic
equation is that for the Schwarzschild case $\omega=K=1$. Following the
standard procedure, we find that the trajectories are ellipses with a perihelia
precession given by 

\begin{equation} \Delta\varphi_p=6\pi{m^2c^2K\over
B^2\omega^3}={6\pi m\over b(1-e^2)}{K\over\omega} 
\label{peri} 
\end{equation}

\noindent where $b$ is the semimajor axis of the ellipse and $e$ is its
eccentricity. Again the difference with the Schwarzschild solution is the
$\omega$ and $K$ multiplying the perihelia precession of the Schwarzschild
solution in (\ref{peri}). In the first approximation in $m$, there is no
difference between equation (\ref{geope}) and the one obtained from the
Schwarzschild solution, since $\omega$ and $K$ depend only on $m^2$. Therefore
there is no difference between (\ref{geope}) and that for the Schwarzschild
solution for the calculation of null geodesics, since this is always made in
the first approximation in $m$. For the calculation of the trajectories of
particles, there is some difference only in the second approximation in $m$,
given by $\omega$ and $K$.  Since $A^2$ must be of order of the energy of the
test particle at infinity, $A^2<<c^2$, the term ${a\over B^2}m^2A^2<<{a\over
B^2}m^2c^2$ for a test particle. So we can set $K=1$ and
$\omega=\sqrt{1+{am^2\over B^2}\epsilon}$ without loss of generality. For a
test particle like Mercure, $\omega^2-1={a\over B^2}m^2c^2={a\over
B^2}{G^2\over c^2}M_{\odot}^2=a\ 2.48\ 10^{-8}$. Here $G$ is the universal
gravitation's constant, $B=2.78\ 10^{15}$ m$^2/$seg is the angular momentum of
Mercure per unity of mass and $M_{\odot}$ is the mass of the sun. This means
that the difference between the Schwarzschild geodesics and the (\ref{geou}) 
geodesics for stars like the sun is then too small to be measured. Let us
assume for a moment that we could take these theories as realistic, then we
conclude that if a star of the size of the sun contains a scalar field inherent
in it, we could not know it because its interaction with the rest of the world
is too small to be detected. Nevertheless, for a pulsar of mass $M=2M_{\odot}$,
which matter is typicaly contained in a radius of $r=10$ Km$\sim 3m$, the
scalar interaction cannot be neglected. Thus, such interactions should be
detectable in stronger gravitational fields like pulsars, where the
gravitational field is much more stronger. 

We have seen that the KK and the LESS theories predict the existence of
magnetic dipoles coupled with gravitational objects naturally, here the
electromagnetic field is a consequence of the natural coupling predicted by the
theory. If we would like to model a pulsar by such a theory, we would not need
to explain the origin of the magnetic dipole in it using internal hypothesis,
since this magnetic dipole would be then a consequence of some more general
interaction between gravitation and electromagnetism.  The price we must pay is
the existence of a scalar field which has not been measured till now. 
Nevertheless, the KK and the LESS theories predict that even if the magnetic
dipole field can be felt around the body, the scalar field interaction is so
weak that it can be measured only near to a distance of order of the
Schwarzschild radius $r_s$. This is so because of the behavior of the scalar
field (see Fig.3) 

\[ 
\Phi={1\over 2\alpha}[\ln(1-{2m\over
r})+\beta\ln({Q\cos\theta\over(r-m)^2-m^2\cos^2\theta}+1)].  
\]


\begin{figure}[h]
\centerline{ \epsfysize=5cm \epsfbox{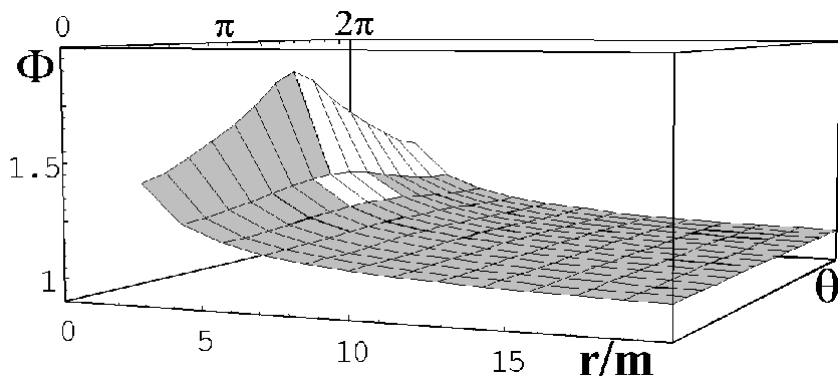}}
\caption{The scalar field (dilaton) for the magnetized Schwarzschild 
solutions. The plot is made in $m$ unities on the horizontal axis.}
\end{figure}

Near to $r_s,\ \Phi$ grows, but it is constant after a few times the value of
$r_s$. Thus the scalar interaction vanishes very rapidly far away from a
distance $r_s=2m$, and it is attractive or repulsive depending on $\alpha$ is
positive or negative. Hence, according to these theories, there exist objects
which posses a fundamental magnetic dipole moment, which is a consequence of a
more general gravito-electromagnetic interaction which posses a scalar field.
Otherwise, according to these theories, even if an astrophysical object like
the sun would posses a scalar field inherent in it, we would not be able to
measure it because of the small force provoked by it. Nevertheless, this
attractive or repulsive scalar force could have effects in stronger
gravitational fields that we should see in astrophysical bodies, but to predict
them, we must solve the geodesic equation (\ref{geo}) near to $r_s$. 

This work is partially supported by CONACYT-Mexico.

\end{document}